\documentclass[twocolumn,aps,prl,superscriptaddress,showpacs]{revtex4}
\usepackage{hyperref}

\usepackage{graphicx}

\usepackage{amsmath}
\usepackage{epsfig}

\begin{document}
 \title{Experimental Demonstration of Continuous Variable Cloning with Phase-Conjugate Inputs}

\author{Metin Sabuncu}
\affiliation{Department of Physics, Technical University of Denmark, Building 309, 2800 Kgs. Lyngby, Denmark}
\affiliation{Institut f\"{u}r Optik, Information und Photonik, Max-Planck Forschungsgruppe, Universit\"{a}t Erlangen-N\"{u}rnberg, G\"{u}nther-Scharowsky str. 1, 91058, Erlangen, Germany}

\author{Ulrik L. Andersen}
\affiliation{Department of Physics, Technical University of Denmark, Building 309, 2800 Kgs. Lyngby, Denmark}
\affiliation{Institut f\"{u}r Optik, Information und Photonik, Max-Planck Forschungsgruppe, Universit\"{a}t Erlangen-N\"{u}rnberg, G\"{u}nther-Scharowsky str. 1, 91058, Erlangen, Germany}

\author{Gerd Leuchs}
\affiliation{Institut f\"{u}r Optik, Information und Photonik, Max-Planck Forschungsgruppe, Universit\"{a}t Erlangen-N\"{u}rnberg, G\"{u}nther-Scharowsky str. 1, 91058, Erlangen, Germany}

\date{\today}

\begin{abstract}
We report the experimental demonstration of continuous variable cloning of phase conjugate coherent states as proposed by Cerf and Iblisdir (Phys. Rev. Lett. 87, 247903 (2001)). In contrast to the proposal of Cerf and Iblisdir, the cloning transformation is accomplished using only linear optical components, homodyne detection and feedforward. Three clones are succesfully produced with fidelities about 89\%. 
\end{abstract}

 \pacs{03.67.Hk, 03.65.Ta, 42.50.Lc}

\maketitle 
One of the most intriguing results of quantum mechanics is that an unknown quantum state cannot be exactly cloned~\cite{wootters82.nat,dieks82.pla}. This fact stems from the inherent linearity of quantum mechanics, and is one of the most discussed features in recent years because it enables secure quantum communication such as quantum key distribution and secret sharing~\cite{scarani05.rev,cerf06}. Because quantum cloning is, in general, imperfect one is led to the construction of optimal, but imperfect, quantum cloning machines. Such machines based on qubits have been experimentally realized in several different settings. On the other hand, cloning machines based on continuous variables have only very recently been implemented~\cite{andersen05.prl,koike06.prl}.   

Yet another interesting feature of quantum mechanics was discovered in 1999 by Gisin and Popescu~\cite{gisin99.prl}. They realized that more quantum information can be encoded into pairs of anti-parallel spins than in parallel ones. The continuous variable (CV) analogue of this effect was addressed by Cerf and Iblisdir who showed that more CV quantum information can be encoded into pairs of phase conjugate coherent states, $|\alpha\rangle |\alpha ^*\rangle$, than in pairs of identical coherent states, $|\alpha\rangle |\alpha\rangle$~\cite{cerf01.pra}. Because of the existence of a strong link between cloning and measurement theory, the superiority of using anti-parallel spins or phase conjugate coherent states led Cerf and Iblisdir to suggest that cloning machines with such inputs perform better than conventional cloning machines. This was indeed the case as shown theoretically in ref. \cite{cerf01.prl} for phase conjugate coherent states and in ref. \cite{fiurasek02.pra} for anti-parallel spin states. Recently, these results were generalized to d-dimensional systems~\cite{zhou06.pra}.


It was also realised by Cerf and Iblisdir that the physical implementation of the cloning of phase conjugate input states is composed of a sequence of beam splitters, a nonlinear process and another sequence of beam splitters~\cite{cerf01.prl}. In this Letter we propose and experimentally realize a much more elegant approach for phase conjugate cloning which is not relying on a non-linear parametric process. A simple combination of beam splitters, detectors and feedforwards suffice to enable optimal N+N$\rightarrow$M Gaussian cloning with phase conjugate input states, where $N$ replicas of $|\alpha\rangle$ and $N$ replicas of $|\alpha ^*\rangle$ serve as inputs to produce M clones. Theoretically, we treat the general case of N+N$\rightarrow$M cloning while the 1+1$\rightarrow$2 and 1+1$\rightarrow$3 cloning of phase conjugate inputs is experimentally demonstrated. We note that the cloning protocol with phase conjugate inputs realized in this Letter have never been implemented before in any quantum system. It is, to the best of our knowledge, the first example of a continuous variable quantum information processing experiment for which there is no experiment with discrete variables.

\begin{figure}[h] \centering \includegraphics[width=8cm]{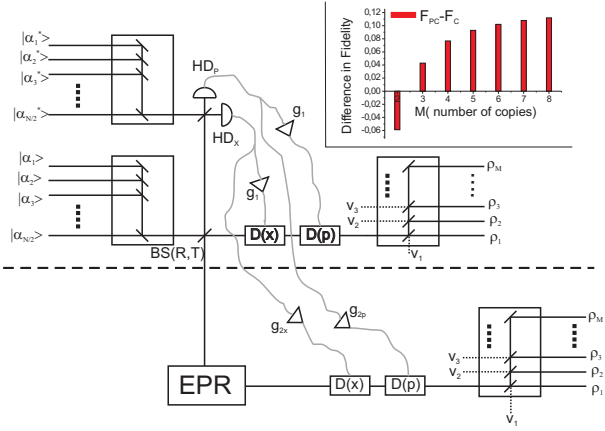} \caption{Proposed setup for an $N+N\rightarrow M$ cloning machine. BS(R,T): Variable beam splitter with transmission T and reflection R. $HD_{x(p)}$: Homodyne detector measuring x(p). $g_1$, $g_{2x}$ and $g_{2p}$: Electronic gains. D(x,p): Displacers of x and p. EPR: Einstein-Podolsky-Rosen entanglement source. $v_i$ denote the vacuum inputs while the $\rho_i$ denote the density operators of the outputs.}
\label{fig1}  \end{figure}

The figure of merit normally used to quantify the quality of a cloning transformation is the average fidelity, which is a measure of the similarity between the input states and the clones. When considering a flat distribution of coherent states as inputs and assuming that the cloning transformation conserves the Gaussian statistics of the quadratures, the optimal 2N$\rightarrow$M cloning machine yields the average fidelity~\cite{cerf00.pra} 
\begin{equation}
F_{C}=\frac{2MN}{2MN+M-2N}
\label{F}
\end{equation}
On the other hand, it has been found that the optimal N+N$\rightarrow$M cloning machine with phase conjugate input states produces clones with fidelity~\cite{cerf01.prl}
\begin{equation}
F_{PC}=\frac{4M^2N}{4M^2N+(M-N)^2}
\label{phaseconj}
\end{equation}
Here $2N$ corresponds to the total number of inputs consisting either of 2N replicas of $|\alpha\rangle$ in Eq. (1) or of N replicas of $|\alpha\rangle$ and N replicas of $|\alpha ^*\rangle$ in Eq. (2). The difference of the two fidelities for N=1 and various numbers of the outputs is depicted in inset of Fig.~\ref{fig1}, and it is clearly seen that the phase conjugate cloning machine outperforms the conventional one for $M>2$. As an example, when a single pair of phase-conjugate inputs (N=1) is transformed into three clones (M=3) the fidelities are $F_{PC}=90\%$ and $F_{C} = 85.7\%$.

A schematic of the cloning machine proposed in this letter is presented in Fig.~\ref{fig1} The input signal is contained in an ensemble consisting of N identical coherent states and N identical phase conjugate coherent states. The two sets of states are uniquely described by the relations $|\alpha\rangle ^{\otimes N} =|x+ip\rangle^{\otimes N}$ and $|\alpha ^*\rangle ^{\otimes N} =|x-ip\rangle^{\otimes N}$, where $x$ is the amplitude quadrature, $p$ is the phase quadrature and $[x,p]=2i$. Each of the two sets of states are then collected into two single states using two arrays of N~-~1 beam splitters. In the Heisenberg picture, the amplitude quadratures after collection can be written as
\begin{eqnarray}
\hat{x}_{c1}=\frac{1}{\sqrt{N}}\sum_{k=1}^{N}\hat{x}_k= \frac{1}{\sqrt{N}}\left( N\langle \hat{x}\rangle+\sum_{k=1}^{N}\delta \hat{x}_k\right)\\
\hat{x}_{c2}=\frac{1}{\sqrt{N}}\sum_{l=1}^{N}\hat{x}'_l= \frac{1}{\sqrt{N}}\left( N\langle \hat{x}\rangle+\sum_{l=1}^{N}\delta \hat{x}'_l\right)
\end{eqnarray}
where $\hat{x}_{k,l}$ have been decomposed as $\hat{x}_{k,l}=\langle \hat{x}\rangle +\delta \hat{x}_{k,l}$ and $\langle \hat{x}\rangle=\langle \hat{x}_k\rangle=\langle \hat{x}_l\rangle$.
The collective coherent state impinges on a beam splitter (with transmission $T$ and reflection $1-T$) and the reflected part is measured jointly with the collective phase conjugated coherent state using a symmetric beam splitter and two homodyne detectors measuring the quadratures $x$ and $p$. The measurement outcomes are electronically amplified with gain $g$ and used to displace the remaining part of the collective coherent state. Such a combination of linear optics, measurements and feedforward enables a shot noise limited amplification if $g_1=\sqrt{2(1-T)/T}$, and the input-output relation is simply~\cite{josse06.prl}
\begin{eqnarray}
\hat{x}_{out}=\sqrt{\frac{1}{T}}\hat{x}_{c1}+\sqrt{\frac{1-T}{T}}\hat{x}_{c2}
\end{eqnarray}
Finally, the resulting state is divided into M clones using an M-splitter. The relation between the inputs and any output clone is thus 
\begin{eqnarray}
\hat{x}_{i}=\frac{1}{\sqrt{TNM}}\Big((N+N\sqrt{1-T}) \langle \hat{x} \rangle + \\\nonumber
\sum_{k=1}^N{\delta \hat{x}_k}+ \sqrt{1-T}\sum_{l=1}^N{\delta \hat{x'}_l}\Big)+
\sum_{j=1}^{M-1}\kappa_{ij}\hat{x}_{vj}\nonumber
\end{eqnarray}
where $\hat{x}_{vj}$ represents vacuum fluctuations and $\kappa_{ij}$ are coefficients that depend on M.
Universal cloning, that is, cloning with conservation of the mean value of $\hat{x}$, is obtained for $T=4MN/(M+N)^2$. In that case the variance of the amplitude quadrature noise of any clone is 
\begin{equation}
\Delta^2 \hat{x}_{i}=1+\frac{(M-N)^2}{2M^2N}.
\end{equation}
The same analysis applies for the phase quadrature and thus it is readily verified that the cloning transformation is symmetric in $\hat{x}$ and $\hat{p}$: $\Delta^2 \hat{p}_{i}=\Delta^2 \hat{x}_{i}$. Now, by using the relation for universal cloning fidelity of coherent states   
\begin{equation}
F=\frac{2}{\sqrt{(1+\Delta^2 \hat{x}_{i})(1+\Delta^2 \hat{p}_{i})}},
\label{fidelity} 
\end{equation}
we immediately retain the optimal fidelity in (\ref{phaseconj}). Hence, optimal Gaussian cloning of phase conjugate coherent states can be obtained using simple linear optics, homodyne detection and feedforward. 

We note that the presented cloning machine is nonunitary. To ensure unitarity, an Einstein-Podolsky-Rosen (EPR) entangled resource must be applied as shown below the dashed line in Fig.~\ref{fig1}: One half of the EPR state is injected into the beam splitter BS(R,T) while the second half is displaced according to the measurement outcomes of the measurements (scaled with gains $g_{2x}=\sqrt{2/T}$ and $g_{2p}=-\sqrt{2/T}$). The latter half is subsequently divided into M clones of $|\alpha ^*\rangle$. Therefore this unitary cloning machine produces not only M optimal coherent state clones but also M optimal phase conjugate coherent state clones. 


\begin{figure}[h] \centering \includegraphics[width=8cm]{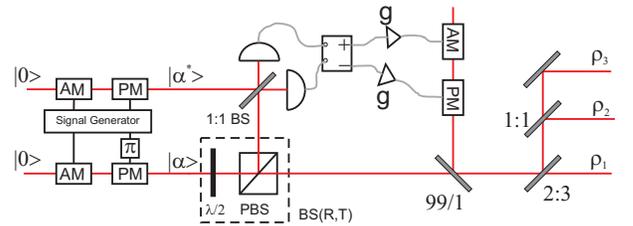} \caption{Schematic of the experimental setup. AM: Amplitude modulator, PM: Phase modulator, g: electronic gains. $\rho_1$, $\rho_2$ and $\rho_3$ are the density operators of the clones.}
\label{fig2}  \end{figure}

We now proceed with the experimental demonstration of the production of 2 and 3 clones from a single pair of phase conjugate coherent states. The experimental setup is shown in Fig.~\ref{fig2}. As a laser source we used a Nd:YAG laser emitting light at 1064nm. The laser beam was split into three parts; two parts served as input modes whereas the last part was used as auxiliary and local oscillator beams. Coherent states are generated at sidebands of bright laser modes. These sidebands, originally in the vacuum state, are excited by the use of amplitude and phase modulators driven by a signal generator. One of the phase modulators is driven $\pi$ out of phase with respect to the other one to ensure the production of phase conjugate beams. In contrast, the amplitude modulators are driven in phase. The phase relation between the two input states is verified by interfering the two states on a 50/50 beam splitter and subsequently measuring the amplitude and phase quadratures in the two outputs of the beam splitter. Extinction of the amplitude (phase) quadrature in the difference (sum) output port of the beam splitter is a clear signature of the preparation of states with proper phase relation and identical amplitudes. 

After preparation of the pair of phase conjugate coherent states ($|\alpha\rangle$ and $|\alpha^*\rangle$), they are injected into the cloning machine. A tunable beam splitter, consisting of a half wave plate and a polarizing beam splitter, separates the coherent state, $|\alpha\rangle$, into two parts. For the production of two or three clones the transmission was set to T~=~8/9 and T~=~3/4, respectively in order to optimise the cloning fidelity. The reflected part of the input state interferes with the phase conjugate state, $|\alpha ^*\rangle$, on a balanced beam splitter. The carrier power of the input modes has been tailored such that the powers of the two states in the joint measurement are balanced. This enables the joint measurements of $x$ and $p$ using the simplified setup shown in Fig.~\ref{fig2} (rather than the standard heterodyne setup consisting of two homodyne detectors): After interference at the beam splitter with a $\pi /2$ relative phase shift, the outputs are directly detected. Subsequently, the sum and difference currents are produced which represent the sum of amplitude quadratures and the difference of the phase quadratures of the two inputs to the beam splitter, respectively. The electronic gains in the feedforward loop are adjusted to ensure close to unity cloning gains. The scaled measurement outcomes are used to modulate an auxiliary beam which subsequently is coupled with the remaining part of $|\alpha\rangle$ using a very asymmetric beam splitter (with splitting ratio 99/1). This accomplishes the displacement operation, and finally the two or three clones are produced by using a single symmetric beam splitter (not shown in Fig.~\ref{fig2}) or two beam splitters with ratios 2:3 and 1:1 respectively.

Knowing that the Wigner functions of the input states and the output clones are Gaussian, they are fully characterized by measuring the first and second order moments of the amplitude and phase quadratures. This is done by using homodyne detection where the local oscillator is stably locked for accessing either the amplitude or phase quadrature. The mean and variance at the sideband were analyzed using a spectrum analyzer which selects the frequency of 14.3 MHz with a resolution bandwidth of 100kHz.

\begin{figure}[h] \centering \includegraphics[width=8cm]{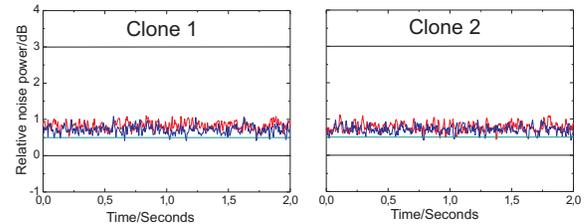} \caption{Spectral noise powers relative to the shot noise for the 2 clones. Red and blue traces correspond to the added cloning noise with respect to the input state for the phase and amplitude quadrature respectively. The 1$\rightarrow$2, $1+1\rightarrow$2 and 2$\rightarrow$2 cloning limits are illustrated by solid straight lines at the 3~dB, 0.5~dB and 0~dB levels. The data was corrected taking into account the homodyne detection efficiencies which were measured to be $83\%$ and  $85\%$  for clone 1 and 2. The noises were measured to $\Delta^2x_1=1.15 \pm 0.2$ and $\Delta^2p_1=1.18 \pm 0.2$ for clone 1, and $\Delta^2x_2=1.19 \pm 0.2$ and $\Delta^2p_2=1.19 \pm 0.2$ for clone 2. The optical gains for this particular measurement run were: $G_{p1}=1,01\pm0,01$, $G_{x1}=1,02\pm0,01$, $G_{p2}=1,00\pm0,01$, and $G_{x2}=0,99\pm0,01$.  Video bandwidth: 30Hz and sweep time: 2 seconds.}
\label{fig3}  \end{figure}

\begin{figure*}[t] \centering \includegraphics[width=15cm]{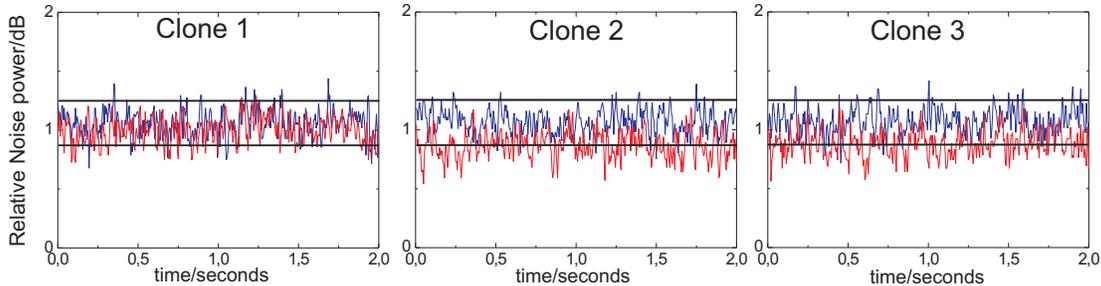} \caption{Spectral noise powers relative to the shot noise for the 3 clones. The straight lines at the 1,25~dB and 0,87~dB level represent the optimal limits for $2\rightarrow3$ and $1+1\rightarrow3$ cloning, respectively. The data was corrected taking into account the homodyne detection efficiencies which were measured to be $83\%$, $85\%$ and $80\%$  for clone 1, 2 and 3. For this particular measurement run we measured the following noises: $\Delta^2x_1=1.26 \pm 0.2$, $\Delta^2p_1=1.27 \pm 0.2$, $\Delta^2x_2=1.22 \pm 0.2$, $\Delta^2p_2=1.28 \pm 0.2$, $\Delta^2x_3=1.23 \pm 0.2$ and $\Delta^2p_3=1.28 \pm 0.2$ and the following gains:  $G_{p1}=1,00\pm0,01$, $G_{x1}=0,95\pm0,01$; $G_{p2}=0,95\pm0,01$, $G_{x2}=0,96\pm0,01$; $G_{p3}=1,01\pm0,01$, and $G_{x3}=1,00\pm0,01$ }
\label{fig4}  \end{figure*}

In order to determine accurately the optical gains for the various clones, we measured the signal power of the input and output. By comparing these results we estimate the gains, the exact values of which can be found in the caption text of Fig.~\ref{fig3} and Fig.~\ref{fig4}. In the following we assume the gains to be unity and consider later implications of nonunity gains. After the gain measurements, the input modulators were switched off in order to precisely measure the cloning noise at 14.3 MHz and thus quantify the cloning performance. Typical examples of measurement runs for the production of 2 and 3 clones are shown in Fig.~\ref{fig3} and Fig.~\ref{fig4} respectively. Here the added cloning noises relative to the shot noise of the input state for the amplitude and phase quadratures are displayed. From these measurements and employing Eq.~(\ref{fidelity}) we can easily determine the cloning fidelities. For the 1+1$\rightarrow$2 cloning operation the copies are produced with fidelities $92.4\%\pm1\%$ and $91.3\%\pm1\%$, whereas the 1+1$\rightarrow$3 cloning machine produces copies with fidelities $88.3\%\pm1\%$, $88.9\%\pm1\%$ and $88.7\%\pm1\%$. 

The latter values demonstrate that the 1+1$\rightarrow$3 cloning machine with phase conjugate inputs operate very close to the optimum of $F_{PC}$=90\% and it outperforms the conventional cloning machine which ideally yields a fidelity of 85.7\% (Eq.~(\ref{F})). The close to optimal performance is a result of the high quality of the feedforward loop. The quantum efficiencies of the detectors (including the interference visibility) were measured to 93\% and the electronic noise was negligible. The 1+1$\rightarrow$2 cloning fidelities are of course not surpassing the fidelity for the conventional 2$\rightarrow$2 cloning machine which trivially yields a fidelity of 100\%. However, using phase conjugate inputs, there exists also the possibility of producing copies of $|\alpha ^*\rangle$ in addition to copies of $|\alpha\rangle$, which is not possible with the conventional cloning approach.  

In the calculations of the fidelities we assumed unity gains. There was however a small deviation from unity, and thus strictly speaking the cloning machine is not universal with respect to the flat coherent state alphabet. The set of coherent input states must be restricted to a certain region in phase space, and the average fidelity for this set of states must be computed. Assuming that the input alphabet is restricted to a Gausssian distribution with variance equal to 10 vaccum units, the average fidelities can be determined to 0.87, 0.87 and 0.89 for the 1+1$\rightarrow$3 cloning machine and 0.92 and 0.92 for the 1+1$\rightarrow$2 cloning machine\cite{cochrane}.


We have now experimentally proved the surprising fact that a cloning machine with phase conjugate input states performs better than a cloning machine with identical inputs for N=1 and M=3. The fact that a pair of phase conjugate coherent states is more informative than identical ones led to a suppression of the noise induced by the cloning action. This close relation between the cloning noise and the information content of the input states is easily understood from the part of the setup executing a joint measurement of phase conjugated coherent states. Such a measurement strategy has recently been proven to be superior for information retrieval of phase conjugate states~\cite{niset06.arc}. In contrast, for identical coherent states such non-local measurement strategy has no advantage over the standard local strategy. Thus the phaseconjugation combined with the joint measurement strategy yields more information which in turn leads to less noise in the displacement operation and subsequently less noise added to the clones.

It is also clear from the setup that the production of infinitely many clones ($M\rightarrow\infty$) coincide with optimal estimation as proved by Bae and Acin~\cite{bae06.prl}. For $M\rightarrow\infty$, the transmission $T\rightarrow 0$ which results in a complete joint measurement of the phase conjugate coherent states, where $|\alpha \rangle$ and $|\alpha ^* \rangle$ interfere at a 1:1 beam splitter and conjugate quadratures are measured. This measurement strategy coincide with the optimal one for estimating the information in phase conjugate coherent states~\cite{niset06.arc}, thus illustrating the strong link between cloning and measurement theory.  

We thank R. Filip for stimulating discussions. This work has been supported by the EU project COVAQIAL (project no. FP6-511004).


\begin{thebibliography}{}



\bibitem{wootters82.nat} W.K. Wootters and W.H. Zurek, {\it Nature} {\bf 299}, 802 (1982). 
\bibitem{dieks82.pla} D. Dieks, {\it Phys. Lett. A} {\bf 92}, 271 (1982).

\bibitem{scarani05.rev} V. Scarani, S. Iblidir, N. Gisin and A. Acin, {\it Rev. Mod. Phys.} {\bf 77}, 1225 (2005). 
\bibitem{cerf06} N.J. Cerf and J. Fiurasek, in: {\it Progress in Optics,} {\bf 49}, edited by E. Wolf, (Elsevire, Amsterdam, 2006), pp. 455.
\bibitem{andersen05.prl} U.L. Andersen, V. Josse, and G. Leuchs, {\it Phys. Rev. Lett.} {\bf 94}, 240503 (2005).
\bibitem{koike06.prl} S. Koike, H. Takahashi, H. Yonezawa, N. Takei, S.L. Braunstein, T. Aoki and A. Furusawa, {\it Phys. Rev. Lett.} {\bf 96}, 060504 (2006).
\bibitem{gisin99.prl} N. Gisin and S. Popescu, Phys. Rev. Lett. {\it 83}, 432 (1999).
\bibitem{cerf01.pra} N. J. Cerf and S. Iblisdir, Phys. Rev. A {\bf 64}, 032307 (2001).
\bibitem{cerf01.prl} N. J. Cerf and S. Iblisdir, Phys. Rev. Lett. {\bf 87}, 247903 (2001).
\bibitem{fiurasek02.pra} J. Fiurasek, S. Iblisdir, S. Massar, and N.J. Cerf, Phys. Rev. A {\bf 65}, 040302 (2002). 
\bibitem{zhou06.pra} X.-F- Zhou, Y.-S. Zhang and G.-C. Guo , Phys. Rev. A {\bf 74}, 042324 (2006). 

\bibitem{cerf00.pra} N. Cerf and S. Iblisdir, {\it Phys. Rev. A} {\bf 62}, 040301(R) (2000).
\bibitem{josse06.prl} V. Josse, M. Sabuncu, N. Cerf, G. Leuchs and U.L. Andersen, Phys. Rev. Lett. {\bf 96}, 163602 (2006).
\bibitem{cochrane} P. T. Cochrane, T. C. Ralph, and A. Doliska, Phys. Rev. A {\bf 69}, 042313 (2004).
\bibitem{niset06.arc} J. Niset, A. Acin, U. L. Andersen, N. J. Cerf, R. Garcia-Patron, M. Navascues, M. Sabuncu, quant-ph/0608215  
\bibitem{bae06.prl} J. Bae and A. Acin, Phys. Rev. Lett. {\bf 97}, 030402 (2006).

\end{thebibliography}
\end{document}